\documentclass[preprint,showpacs,showkeys,preprintnumbers,amsmath,amssymb]{revtex4}
 \usepackage{dcolumn}
 \usepackage{bm}

\begin{document}

 \title{ Dirac particles tunneling from BTZ black hole }

 \author{Ran Li}
 \thanks{Corresponding author. Electronic mail: liran05@lzu.cn}
 \author{Ji-Rong Ren}
 \affiliation{Institute of Theoretical Physics, Lanzhou University, Lanzhou, 730000, Gansu, China}

 \begin{abstract}

 We calculated the Dirac Particles' Hawking radiation
 from the outer horizon of BTZ black hole via tunneling
 formalism. Applying WKB approximation to the Dirac
 equation in (2+1) dimensional BTZ spacetime background,
 we obtain the radiation spectrum for fermions and Hawking temperature of
 BTZ black hole. The results obtained by taking the fermion
 tunneling into account is consistent with the previous literatures.

 \end{abstract}

 \pacs{04.70.Dy, 03.65.Sq}

 \keywords{tunneling, Hawking radiation, BTZ black hole}

 \maketitle

 Hawking\cite{hawking} discovered the thermal radiation of a collapsing black hole
 using the techniques of quantum field theory in curved
 spacetime. Since the Hawking radiation relates the
 theory of general relativity with quantum
 field theory and statistical thermodynamics, it is generally
 believed that a deeper understanding of Hawking radiation may shed
 some lights on seeking the underlying quantum gravity.
 Since then, several derivations of Hawking radiation have been
 proposed. The original method presented by Hawking is direct but
 complicated to be generalized to other spacetime backgrounds. In
 recent years, a semi-classical derivation of Hawking radiation as
 a tunneling process\cite{parikh} has been developed and has
 already attracted a lot of
 attention. In this method, the imaginary part of the action is
 calculated using the null geodesic equation. Zhang and Zhao extended this method to
 Reissner-Nordstr\"{o}m black hole\cite{zhangjhep} and Kerr-Newman black hole\cite{zhangplb}.
 M. Angheben \textit{et al} \cite{angheben} also proposed a
 derivation of Hawking radiation by calculating the particles'
 classical action from the Hamilton-Jacobi equation, which is an
 extension of the complex path analysis of T. Padmanabhan \textit{et
 al} \cite{padmanabhan}. Both of these approaches to tunneling
 used the fact that the tunneling probability for the classically
 forbidden trajectory from inside to outside the horizon is given
 by
 \begin{eqnarray}
 \Gamma=\textrm{exp}(-\frac{2}{\hbar}\textrm{Im}I)\;.
 \end{eqnarray}
 where $I$ is the classical action of the trajectory.
 The crucial thing in tunneling formalism is to calculate the
 imaginary part of classical action. The difference between these
 two methods consists in how the classical action is calculated.
 For a detailed comparison of the Hamilton-Jacobi method and the
 Null-Geodesic method, one can see \cite{kerner}.

 In this letter, we extend the tunneling method presented in \cite{kernerarxiv}
 to calculate the Dirac particles' Hawking radiation from (2+1) dimensional BTZ black hole.
 Although many work\cite{medved,vagenas,liu,wu,zhao,jiang,he}
 have been contributed to the Hawking radiation
 of BTZ black hole using the tunneling method, they all considered
 the scalar particles' radiation. Starting with the covariant Dirac
 equation in curved background, we calculate the radiation
 spectrum and Hawking temperature by using the WKB approximation.

 The BTZ black hole solution is an exact
 solution to Einstein field equation in a $(2+1)$ dimensional
 theory of gravity with a negative cosmological
 constant $\Lambda=-1/l^2$:
 \begin{equation}
 S=\int dx^3 \sqrt{-g}(R+2\Lambda)\;.
 \end{equation}
 The BTZ black hole is described by the metric\cite{BTZ}
 \begin{equation}
 ds^2=-N^2
 dt^2+\frac{1}{N^2}dr^2+r^2(N^{\phi}dt+d\phi)^2\;,
 \end{equation}
 where
 \begin{equation}
 N^2=-M+\frac{r^2}{l^2}+\frac{J^2}{4r^2}\;,\;\;N^{\phi}=-\frac{J}{2r^2}\;,
 \end{equation}
 with $M$ and $J$ being the ADM mass and angular momentum
 of the BTZ black hole, respectively. This metric is stationary
 and axially symmetric, with two killing vectors $(\partial/\partial
 t)^{\mu}$ and $(\partial/\partial\phi)^{\mu}$. The line element
 (3) has two horizons which is determined by the equation
 \begin{eqnarray}
 N^2=\frac{1}{l^2r^2}(r^2-r_+^2)(r^2-r_-^2)=0\;,
 \end{eqnarray}
 where $r_+$ and $r_-$ are defined by
 \begin{eqnarray}
 r_{\pm}^2=\frac{Ml^2}{2}\big[1\pm\sqrt{1-\frac{J^2}{M^2l^2}}\big]\;.
 \end{eqnarray}
 We have assumed the non-extremal condition $Ml>J$, so that $r_+$ and $r_-$
 correspond to the outer event horizon and the inner event horizon
 respectively which is similar to the Reissner-Nordstr\"{o}m
 spacetime. In the extremal case $Ml=J$, the two horizons
 coincide. In the present paper, we mainly consider the non-extremal
 case. A brief comment regarding the extremal case appears at the end of this paper.

 Now we calculate the Dirac particles' Hawking radiation from the
 BTZ black hole.
 We consider the two component massive spinor field $\Psi$,
 with mass $\mu$, obeys the covariant Dirac equation
 \begin{eqnarray}
 i\hbar\gamma^ae_a^\mu\nabla_\mu\Psi-\mu\Psi=0\;,
 \end{eqnarray}
 where $\nabla_\mu$ is the spinor covariant derivative defined by
 $\nabla_\mu=\partial_\mu+\frac{1}{4}\omega_\mu^{ab}\gamma_{[a}\gamma_{b]}$,
 and $\omega_\mu^{ab}$ is the spin connection, which can be given
 in terms of the tetrad $e_a^\mu$. The $\gamma$ matrices in three
 spacetime dimensions is selected to be $\gamma^a=(i\sigma^2, \sigma^1,
 \sigma^3)$, where the matrices $\sigma^k$ are the Pauli matrices.
 According to the line element (3), the tetrad field $e_a^\mu$ can
 be selected to be
 \begin{eqnarray}
 e_0^\mu&=&(\frac{1}{N}, 0, -\frac{N^\phi}{N})\;,\nonumber\\
 e_1^\mu&=&(0, N, 0)\;,\nonumber\\
 e_2^\mu&=&(0, 0, \frac{1}{r})\;.
 \end{eqnarray}

 We use the ansatz for the two component spinor $\Psi$ as following
 \begin{eqnarray}
 \Psi=\left(%
 \begin{array}{c}
  A(t,r,\phi) \\
  B(t,r,\phi) \\
 \end{array}%
 \right)\textrm{exp}\big[\frac{i}{\hbar}I(t,r,\phi)\big]\;.
 \end{eqnarray}
 In order to apply WKB approximation, we can insert the ansatz
 for spinor field $\Psi$ into the Dirac equation. Dividing by the
 exponential term and neglecting the terms with $\hbar$, one can
 arrive at the following two equations
 \begin{eqnarray}
 \left\{
  \begin{array}{ll}
    A(\mu+\frac{1}{r}\partial_\phi I)+
 B(N\partial_r I+\frac{1}{N}\partial_t I-\frac{N^\phi}{N}\partial_\phi
 I)=0\;,  \\
    A(N\partial_r I-\frac{1}{N}\partial_t I+\frac{N^\phi}{N}\partial_\phi I)
 +B(\mu-\frac{1}{r}\partial_\phi I)=0\;.
  \end{array}
 \right.
 \end{eqnarray}
 Note that although $A$ and $B$ are not constant, their
 derivatives and the components $\omega_\mu$ are all of the factor
 $\hbar$, so can be neglected to the lowest order in WKB
 approximation.
 These two equations have a non-trivial solution for $A$ and $B$ if and only if the
 determinant of the coefficient matrix vanishes. Then we can get
 \begin{eqnarray}
 N^2(\partial_r I)^2-\frac{1}{N^2}(\partial_t I-N^\phi\partial_\phi
 I)^2+\frac{1}{r^2}(\partial_\phi I)^2-\mu^2=0\;.
 \end{eqnarray}
 Because there are two killing vectors $(\partial/\partial
 t)^{\mu}$ and $(\partial/\partial\phi)^{\mu}$ in the BTZ
 spacetime, we can separate the variables for $I(t,r,\phi)$ as
 following
 \begin{eqnarray}
 I=-\omega t+j\phi+R(r)+K\;,
 \end{eqnarray}
 where $\omega$ and $j$ are Dirac particle's energy and angular
 momentum respectively, and $K$ is a complex constant. Insert
 it to equation (11) and solving for
 $R(r)$ yields
 \begin{eqnarray}
 R_\pm(r)=\pm\int\frac{dr}{N^2}\sqrt{(\omega+jN^\phi)^2+N^2(\mu^2-\frac{j^2}{r^2})}\;.
 \end{eqnarray}
 As discussed in the Hamilton-Jacobi method\cite{akhmedov,mitra}, one solution
 corresponds Dirac particles moving away from the outer event horizon and the
 other solution corresponds the particles moving toward the outer event
 horizon. The probabilities of crossing the outer horizon each way are
 respectively given by
 \begin{eqnarray}
 P_{out}&=&\textrm{exp}[-\frac{2}{\hbar}\textrm{Im}I]
 =\textrm{exp}[-\frac{2}{\hbar}(\textrm{Im}R_++\textrm{Im}K)]\;,\nonumber\\
 P_{in}&=&\textrm{exp}[-\frac{2}{\hbar}\textrm{Im}I]
 =\textrm{exp}[-\frac{2}{\hbar}(\textrm{Im}R_-+\textrm{Im}K)]\;.
 \end{eqnarray}
 To ensure that the probability is normalized, we should note
 that the probability of any incoming classical particles crossing
 the outer horizon is unity\cite{mitra}. So we get $\textrm{Im}K=-\textrm{Im}R_-$.
 Since $\textrm{Im}R_+=-\textrm{Im}R_-$ this implies that the
 probability of a particle tunneling from inside to outside the
 outer horizon is given by
 \begin{eqnarray}
 \Gamma=\textrm{exp}[-\frac{4}{\hbar}\textrm{Im}R_+]\;.
 \end{eqnarray}
 The imaginary part of $R_+$ can be calculated using equation
 (13). Integrating the pole at the horizon leads to the result
 (see \cite{kerner,mitra} for a detailed similar process)
 \begin{eqnarray}
 \textrm{Im}R_+=\frac{\pi}{2\kappa}(\omega-\omega_0)\;,
 \end{eqnarray}
 where $\kappa=(r_+^2-r_-^2)/(l^2r_+)$ is the surface gravity of
 outer event horizon and $\omega_0=j\Omega_+$ with
 $\Omega_+=J/(2r_+^2)$ is the angular velocity of the outer event
 horizon. This leads to the tunneling probability
 \begin{eqnarray}
 \Gamma=\textrm{exp}\big[-\frac{2\pi}{\kappa}(\omega-\omega_0)]\;,
 \end{eqnarray}
 which is consistent with the
 previous literatures(a recent discussion appeared in Ref.\cite{he}).
 It should be noted that the higher terms about $\omega$ and $j$
 are neglected in our derivation and the expression (17) for tunneling
 probability implies the pure thermal radiation.
 The higher terms of the tunneling probability can arise from
 the energy and angular momentum conservation.
 In \cite{he}, the authors obtained the emission rate $\Gamma=e^{\Delta S_BH}$
 when taking the back reaction into account and argued the result
 offers a possible mechanism to explain the information loss paradox.

 From the emission probability (17), the fermionic
 spectrum of Hawking radiation of
 Dirac particles from the BTZ black hole
 can be deduced following the standard arguments\cite{damour,sannan}
 \begin{eqnarray}
 N(\omega, j)=\frac{1}{e^{2\pi(\omega-\omega_0)/\kappa}+1}\;.
 \end{eqnarray}
 From the tunneling probability and radiant spectrum, Hawking temperature
 of BTZ black hole can be determined as
 \begin{eqnarray}
 T=\frac{\kappa}{2\pi}=\frac{1}{2\pi l^2 r_+}(r_+^2-r_-^2)\;.
 \end{eqnarray}

 At last, we present a brief comment regarding
 the extremal case $Ml=J$. In the extremal case,
 the two horizons coincide, \textit{i.e.} $r_+=r_-$.
 The factor $N^2(r)$
 is of order two in terms of power series
 expansion near the horizon, namely $N^2(r)\sim (r-r_+)^2$,
 which is very different from the non-extremal case where
 $N^2(r)\sim(r-r_+)$. Then the integral (13) is divergent.
 This yield a diverging real component in the action while
 no imaginary part presented, which
 implies that the extremal black hole can not emit particle
 in order to avoid the creation of naked singularity.
 Then we argue that Hawking temperature for extremal
 BTZ black hole is just zero. The result for the extremal
 black hole don't violate the cosmic censorship.

 In summary, we have calculated the Dirac particles' Hawking
 radiation from BTZ black hole using the tunneling formalism.
 Starting with Dirac equation, we obtained the radiation spectrum
 and Hawking temperature of BTZ black hole by using
 the WKB approximation. The results is
 coincide with the previous literatures. If the method presented
 in this paper is also valid for other background spacetime is
 interesting to investigate in the future.

\section*{ACKNOWLEDGEMENT}

 This work was supported by the National Natural Science Foundation
 of China and Cuiying Project of Lanzhou University.

 \end{document}